\documentclass[3p,times,procedia]{elsarticle}
\usepackage{nupha_ecrc}
\usepackage[utf8]{inputenc}

\volume{00}

\firstpage{1}

\journalname{Nuclear Physics A}

\runauth{}

\jid{nupha}

\jnltitlelogo{Nuclear Physics A}

\usepackage{amssymb}
\usepackage{lineno}

\usepackage[figuresright]{rotating}

\newcommand{\mee}{\ensuremath{m_{\rm ee}}}

\newcommand{\ptee}{\ensuremath{p_{\rm T,ee}}}
\newcommand{\GeVcc}{GeV/$c^2$}
\newcommand{\GeVc}{GeV/$c$}
\newcommand{\MeVc}{MeV/$c$}
\newcommand{\sNN}{\ensuremath{\sqrt{s}_{\rm NN}}}
\newcommand{\dNdmee}{\ensuremath{ {\rm d} N/ {\rm d} \mee} }

\begin{document}

\begin{frontmatter}



%% Instructions from Editor: Please use the following \dochead only in the preprint version (e-print arXiv etc.); 
%% use empty \dochead{} when submitting to Nuclear Physics A!
\dochead{XXVIIIth International Conference on Ultrarelativistic Nucleus-Nucleus Collisions\\ (Quark Matter 2019)}
%% e.g. \dochead{17th International Conference on Dynamical Processes in Excited States of Solids}

\title{Low-mass dielectron measurements in pp, p--Pb, and Pb--Pb collisions with ALICE at the LHC}


\author{H. Sebastian Scheid for the ALICE Collaboration}

% \address{s.scheid@cern.ch \\ Goethe University Frankfurt, Max-von-Laue Straße 1, 60438 Frankfurt}
\address{Goethe University, Frankfurt am Main, Germany}

\begin{abstract}

Dielectrons are an excellent probe for the QCD matter created in created in ultra-relativistic heavy-ion collisions, since they are emitted during the whole evolution of the collision and do not interact strongly with the medium. To isolate the QGP signals, measurement of the dielectron production in vacuum and its modifications due to the presence of cold nuclear matter is necessary.
We present and discuss results from a low magnetic field detector setup in proton-proton collisions at $\sqrt{s} = 13$~TeV, as well as  the measurement of dielectron production in pp , p–Pb, and Pb–Pb collisions at $\sqrt{s_{\rm NN}} = 5$~TeV.

\end{abstract}


\end{frontmatter}


%\linenumbers

\section{Introduction}

In heavy-ion collisions dielectrons originate from various sources, i.~e.~decays of pseudo-scalar and vector mesons, semi-leptonic decays of correlated open heavy-flavour hadrons, and thermal radiation from a Quark--Gluon Plasma (QGP) as well as a hot hadron gas. The latter allows the study of the average temperature of the fireball, and the onset of chiral symmetry restoration. Investigation of these non-trivial phenomena demands a precise understanding of the production of dielectrons from hadronic decays and their initial or final state modifications.
During the second data taking period of the LHC from 2015 to 2018, ALICE collected data of proton-proton (pp), p--Pb, and Pb--Pb collisions at $\sqrt{s_{\rm NN}} = 5.02$~TeV providing the possibility to compare the dielectron production in all systems at the same collision energy per nucleon pair.

In special data taking campaigns data in pp collisions at $\sqrt{s} = 13$~TeV were recorded, while the magnetic field in the central barrel was reduced from 0.5 to  0.2~T, extending the reach of particle measurements to a transverse momentum $p_{\rm T}$ $> 75$~\MeVc. This setup gives ALICE the possibility to measure dielectrons at LHC energies in an unexplored kinematic regime, at low invariant mass and pair transverse momentum.

\section{Results}

\begin{figure}
    \centering
        \includegraphics[scale=0.11]{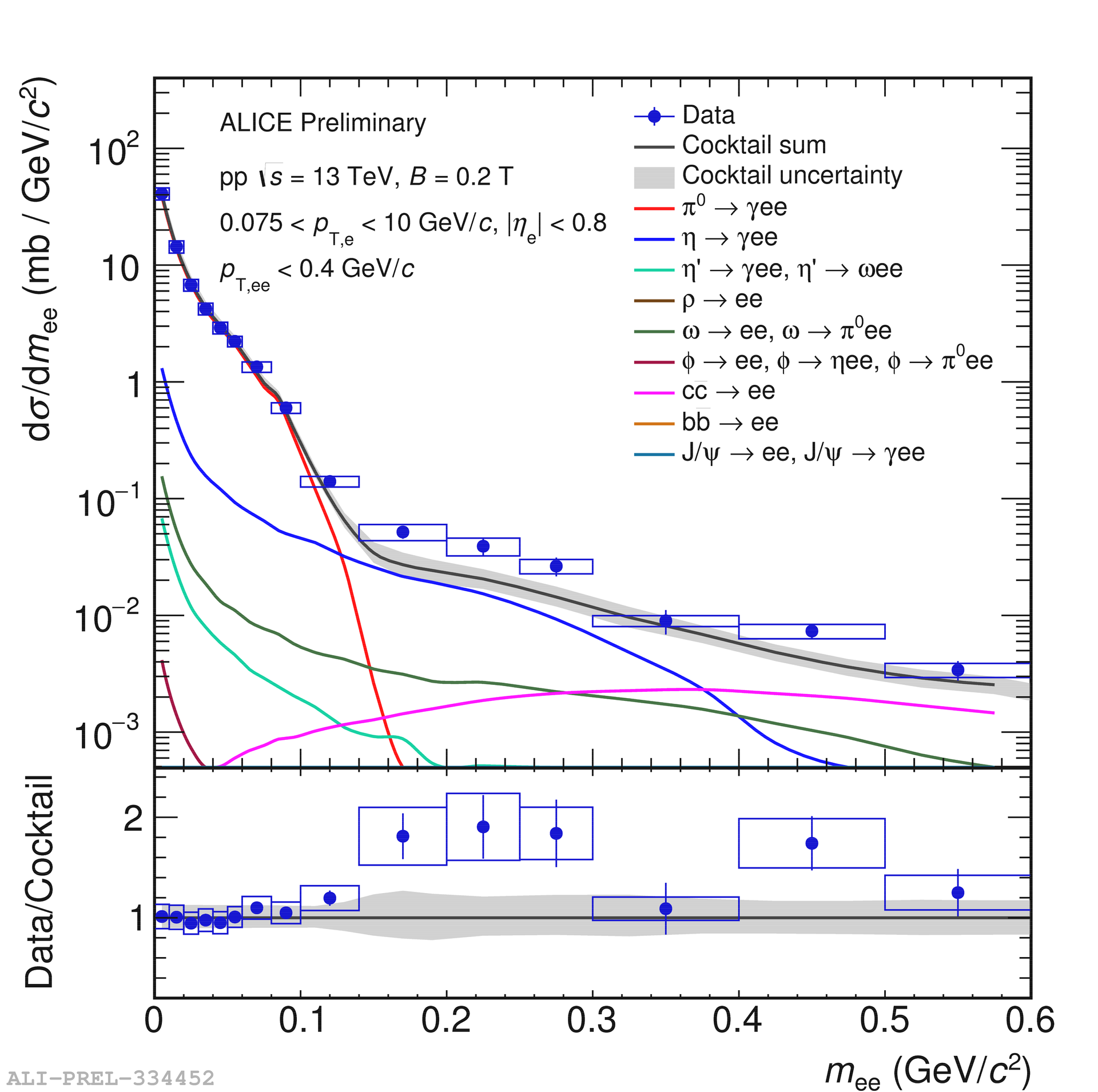}
        \includegraphics[scale=0.11]{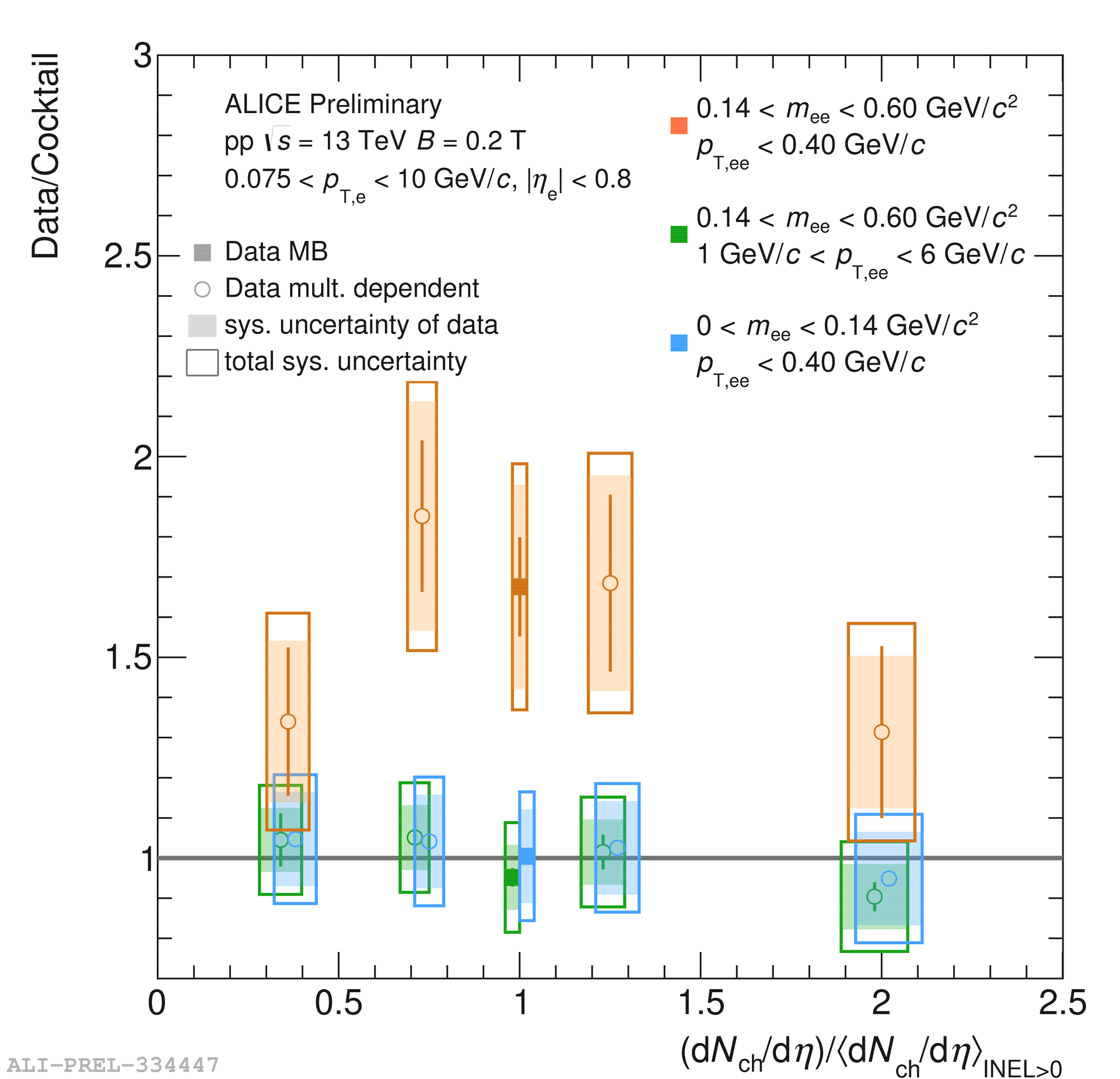}
    \caption{Cross section of dielectron production for $\ptee < 0.4$~\GeVc\ measured in minimum-bias pp collisions at $\sqrt{s} = 13$~TeV with a low magnetic field detector setup compared to expected contributions from hadronic sources (left) and data over cocktail ratios for different kinematic regions as a function of the relative charged particle density (right).}
    \label{fig:lowfield}
\end{figure}

In Fig. \ref{fig:lowfield} (left) the cross section of dielectron production as a function of invariant mass (\mee) is shown in inelastic pp collisions at $\sqrt{s} = 13$~TeV recorded at $B = 0.2$~T.
The  mass spectrum is shown for small pair transverse momentum (\ptee) $< 0.4$~\GeVc\ . The measurement is compared with the expected  contribution from known hadronic sources, the so-called hadronic cocktail, constructed from independent measurements of hadron yields using established methodology~\cite{pp7TeV,pp13TeV}. In the mass region dominated by Dalitz decays of $\pi^{0}$ ($\mee < 0.14$~\GeVcc) the hadronic cocktail is well in agreement with the data, whereas at larger masses an enhancement over the expected hadronic contributions can be observed. The enhancement has a significance of 2.1$\sigma$ taking the statistic and systematic uncertainties into account. Furthermore, Fig. \ref{fig:lowfield} (right) shows the multiplicity dependence of the data to cocktail ratios in different kinematic regions. The data exceeds the expectations from hadronic sources for the kinematic region of $0.14 < \mee < 0.6$~\GeVcc\ ($\eta$ mass region) and $\ptee < 0.4$~\GeVc\ in all multiplicity intervals, where the scaling of the enhancement with the multiplicity is compatible with a linear scaling. In the $\pi^{0}$ mass region in the same \ptee\ range or the $\rm \eta$ mass region at higher \ptee\ ($1 - 6$~\GeVc) the cocktail is in agreement with the data within uncertainties.
The main systematic uncertainty of the cocktail in the mass region of interest could be reduced drastically by a measurement of the $\eta$ at low $p_{\rm T}$.
Previous measurements performed in pp collisions at lower collision energies observed an excess of soft photons and dielectrons in a similar kinematic range~\cite{Baum}, that was discussed in the context of hadronic Bremsstrahlung~\cite{Low}, soft parton annihilation in a cold QGP~\cite{Lichard}, and quark synchrotron radiation~\cite{Nachtmann}.

\begin{figure}
    \centering
        \includegraphics[scale=0.375]{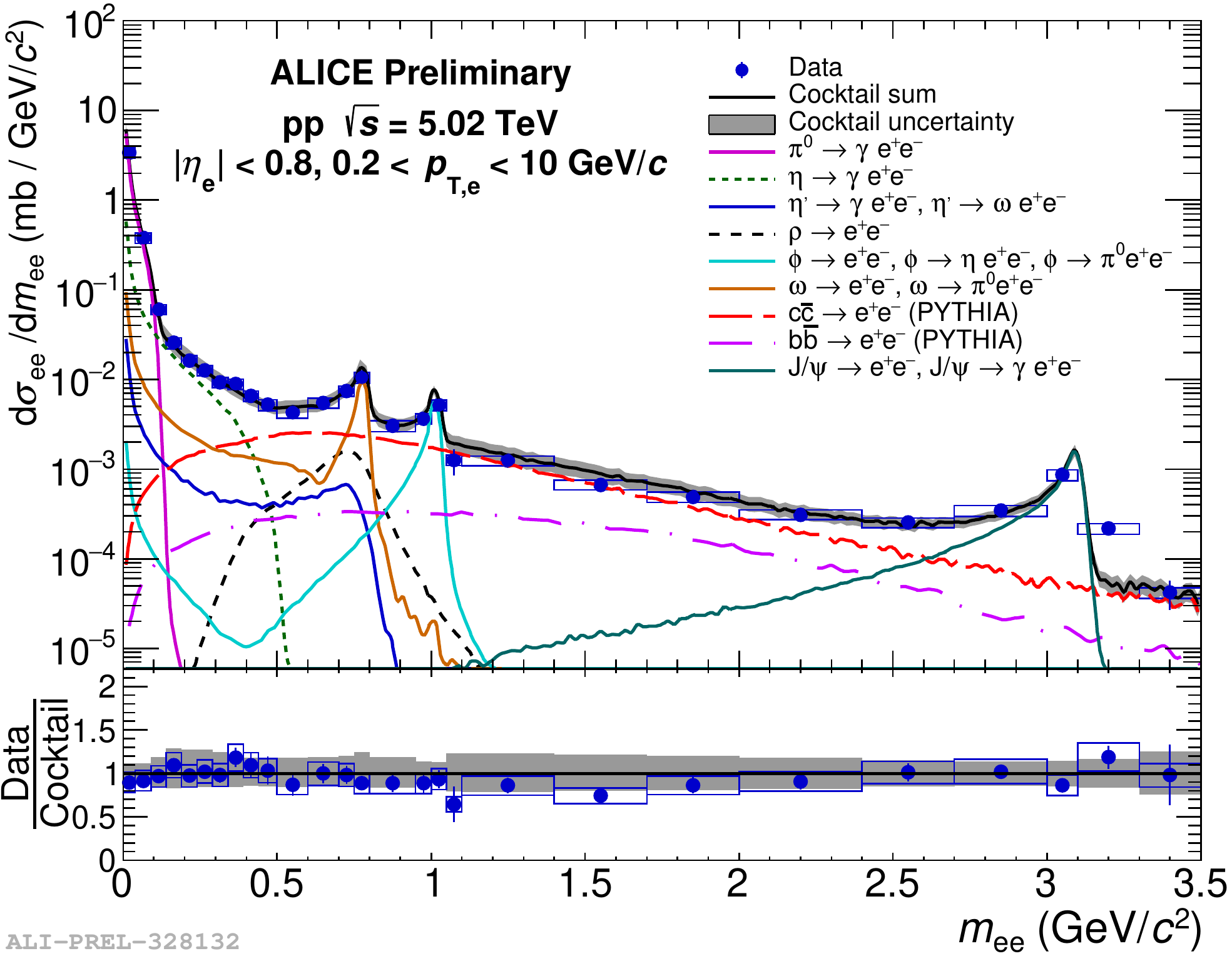}
        \includegraphics[scale=0.105]{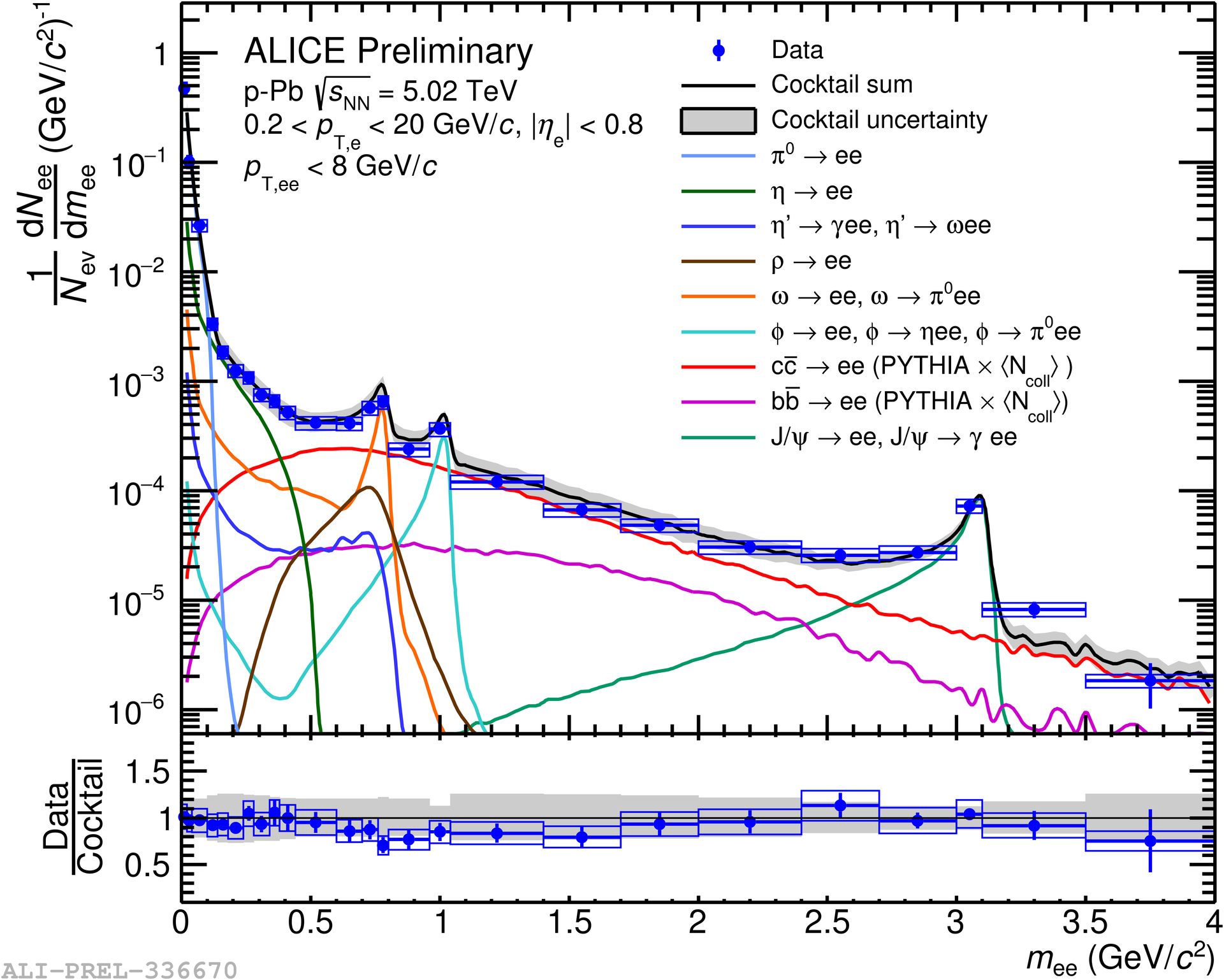}
    \caption{Cross section of dielectron production in pp collisions (left) and dielectron yield in p--Pb collisions (right) as a function of invariant mass measured at $\sqrt{s_{\rm NN}} = 5.02$~TeV.}
    \label{fig:meeSpectra}
\end{figure}

The cross section of dielectron production as a function of \mee\ measured in pp collisions at $\sqrt{s} = 5.02$~TeV is presented and compared to a hadronic cocktail in Fig. \ref{fig:meeSpectra} (left).
The first measurement of the charm ($\rm c\overline{c}$) and beauty ($\rm b\overline{b}$) production cross sections at mid-rapidity in pp collisions at $\sqrt{s} = 5.02$~TeV is performed by fitting templates to the \mee\ and \ptee\ distributions in the intermediate mass region (IMR, $1.1 < \mee < 2.7$~\GeVcc), providing a complementary method to the measurement of the full reconstruction of hadronic decays of heavy-flavour hadrons~\cite{hfCC,Jpsi}. The templates are based on the PYTHIA6~\cite{pythia6} Perugia2011\cite{tune} and POWHEG~\cite{powheg} event generators. The results are summarised in Tab. \ref{tab:hf}. They fall in line with previous measurements at $\sqrt{s} = 7$~\cite{pp7TeV} and 13~TeV~\cite{pp13TeV}, showing the sensitivity of the dielectron measurements to the different implementation of the heavy-flavour production in the model calculations. The energy dependence follows the trend predicted by FONLL~\cite{FONLL} calculations.

\begin{table}[h]
    \centering
    \caption{Cross section of heavy-flavour production at mid-rapidity in pp collisions at $\sqrt{s} = 5.02$~TeV. Systematic uncertainties on effective branching ratios, 22\% for $\rm c\overline{c}\rightarrow e^{+}e^{-}$ and 6\% for $\rm b\overline{b}\rightarrow e^{+}e^{-}$, are not included.}
    \begin{tabular}{c|c|c}
        & PYTHIA & POWHEG \\
        \hline
        \hline
        $ {\rm d}\sigma_{\rm c\overline{c}}/{\rm d}y|_{y=0} $ (mb) & 0.531 $\pm$ 0.062 (stat) $\pm$ 0.066 (syst) & 0.743 $\pm$ 0.080 (stat) $\pm$ 0.093 (syst)\\
        $ {\rm d}\sigma_{\rm b\overline{b}}/{\rm d}y|_{y=0} $ (mb) & 0.037 $\pm$ 0.005 (stat) $\pm$ 0.002 (syst) & 0.027 $\pm$ 0.004 (stat) $\pm$ 0.003 (syst)
    \end{tabular}
    
    \label{tab:hf}
\end{table}

The dielectron yield measured in p--Pb collisions at $\sNN = 5.02$~TeV is depicted in Fig. \ref{fig:meeSpectra} (right).
The data are compared to a hadronic cocktail constructed in the same way as in the pp analysis. The heavy-flavour contributions use the same cross sections as the ones used for the default hadronic cocktail in pp collisions at 5.02~TeV shown in Fig. \ref{fig:meeSpectra} (left), scaled with the number of binary collisions $\langle N_{\rm coll} \rangle = 6.7\pm0.2$~\cite{ncoll}, and taking the asymmetry energy of the colliding beams into account. The cocktail is well in agreement with the data within uncertainties.

\begin{figure}
    \centering
        \includegraphics[scale=0.365]{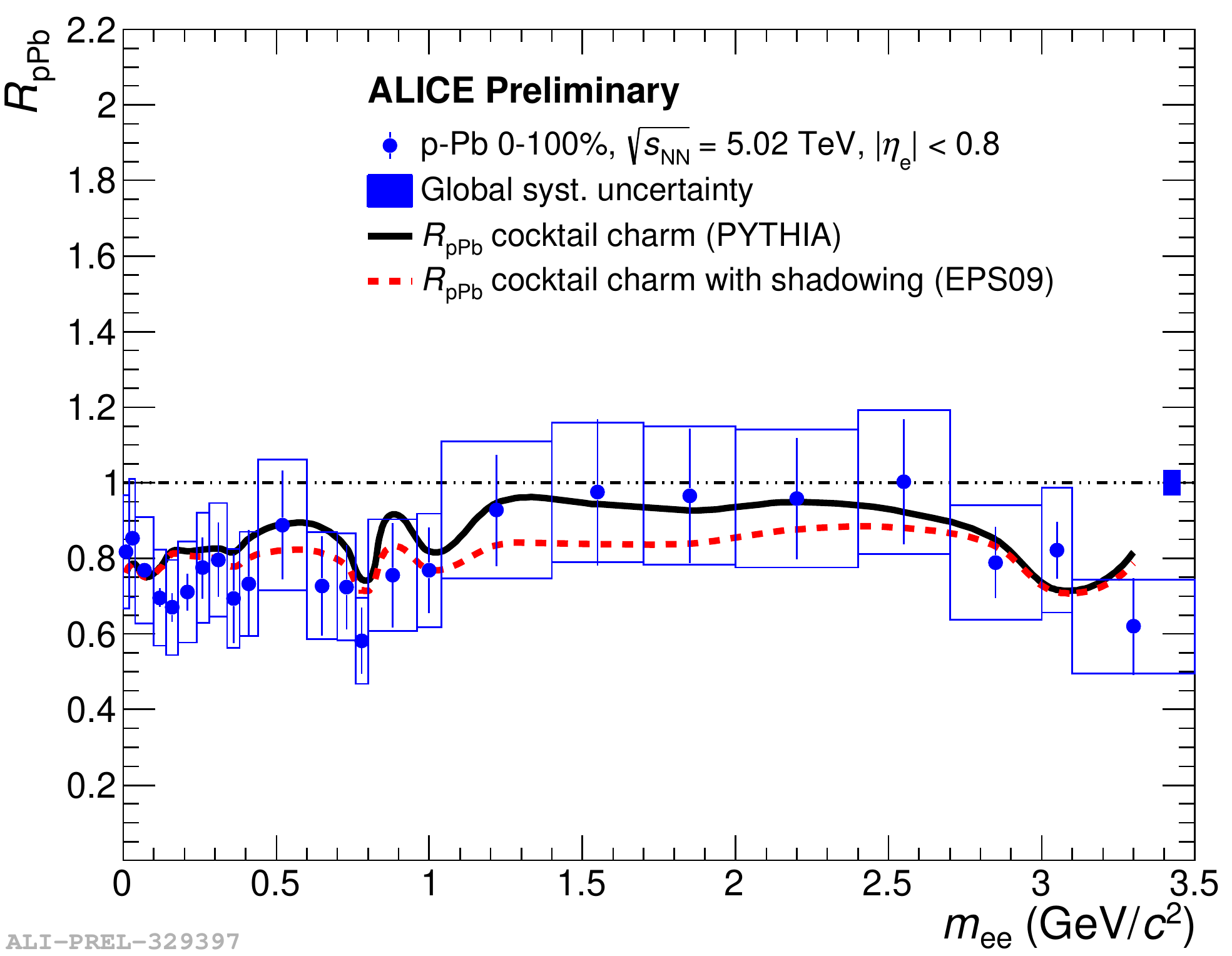}
        \includegraphics[scale=0.365]{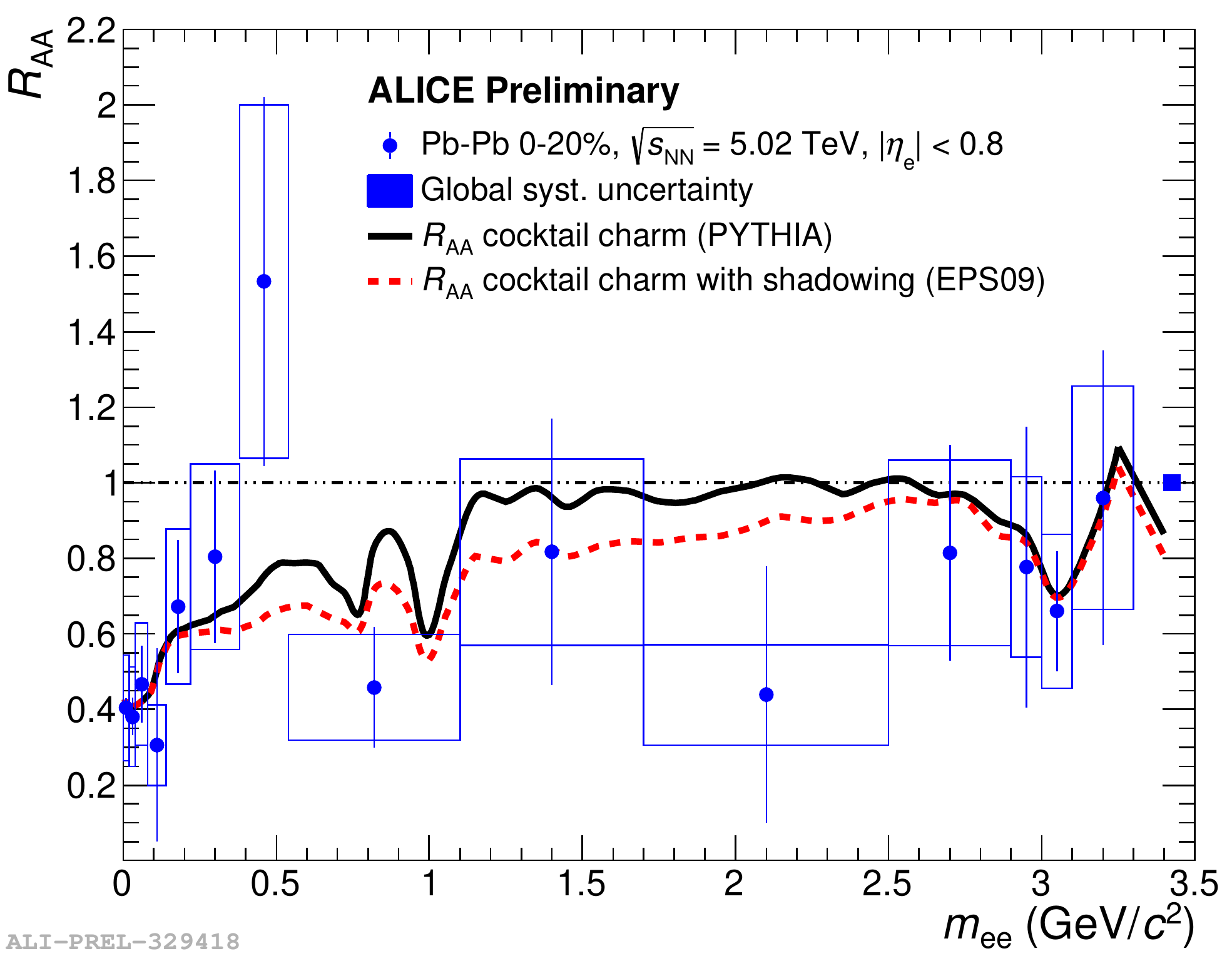}
    \caption{Nuclear-modification factors of dielectrons $R_{\rm pPb}$ from minimum bias p--Pb collisions (left) and $R_{\rm AA}$ from 0-20\% most central Pb--Pb collisions (right) as a function of $m_{\rm ee}$ measured at $\sqrt{s_{\rm NN}} = 5.02$~TeV. Both are compared to expectations from the hadronic cocktail in vacuum (solid line), and including nuclear modifications of the charm contribution from EPS09~\cite{eps09} (dashed line).}
    \label{fig:ratios}
\end{figure}

The nuclear modification factor $R_{\rm pPb}$ has been calculated as $R_{\rm pPb} = \frac{1}{\langle N_{\rm coll} \rangle}\frac{\dNdmee|_{\rm pPb}}{\dNdmee|_{\rm pp}}$, and is  shown in Fig. \ref{fig:ratios} (left) together with two cocktail calculations based on PYTHIA. One cocktail considers $\langle N_{\rm coll} \rangle$ scaled vacuum production of the charm contribution (solid line), whereas a second cocktail includes modifications of the charm production via the nuclear parton distribution functions (nPDFs) from EPS09~\cite{eps09}.
In the IMR the data is consistent with unity, suggesting no modification of the charm production beyond a scaling with $\langle N_{\rm coll} \rangle$ within uncertainties. 
For $\mee < 1.1$~\GeVcc\ the deviation of the data from  unity is expected, since the light flavour sources have been shown to not scale with $\langle N_{\rm coll} \rangle$ in previous measurements. 
Both cocktails can describe the data within the uncertainties.
For $\mee < 1$~\GeVcc\ the EPS09 cocktail is closer to the central value of the data.
% The data are well reproduced by the vacuum cocktail, which suggests no further modifications of the heavy-flavour production in p--Pb with respect to pp collisions beyond a scaling with $N_{\rm coll}$. 
% The cocktail including EPS09 is in agreement with the data within the large uncertainties.
This suggests that, if there is a significant modification of charm production, the sensitive region for dielectron measurements would be in a mass window of about $0.5 < \mee < 1.0$~\GeVcc.

Figure \ref{fig:ratios} (right) shows the $R_{\rm AA}$ constructed from the same pp baseline, together with measurements in 0-20\% most central Pb--Pb collisions at \sNN = 5.02~TeV~\cite{caliva}. In the IMR, a hint for a suppression below the vacuum expectation is observed. Following previous arguments this should not be considered as a modification of the initial state, but rather final state effects, e.g. energy loss or rescattering in a hot and dense medium, which are not included in the cocktail calculations. The precision of the data prohibits a measurement of thermal radiation from a hot hadron gas or QGP that would manifest itself as an enhancement above the hadronic cocktail. 
% At a mass of about 0.5~\GeVcc\ a hint for an enhancement can be observed above the cocktail. 
This enhancement would be more pronounced, taking final state effects on the charm contribution into account.

\section{Summary and Conclusion}
ALICE measured dielectron production in inelastic pp collisions at $\sqrt{s} = 13$~TeV with the magnetic field in the central barrel reduced to 0.2~T which allows the investigation of very soft dielectron production. An excess over the expectation of hadronic sources with a significance of 2.1$\sigma$ is found for $\ptee < 0.4$~\GeVc\ in the \mee window of $0.14 - 0.6$~\GeVcc. The scaling of this excess with the event multiplicity is compatible with a linear behaviour. In other \mee\ or \ptee\ regions the hadronic cocktail is in agreement with the measurements within uncertainties.
The measurement of dielectron production in pp collisions at $\sqrt{s} = 5.02$~TeV yields the first measurements of ${\rm d}\sigma_{\rm c\overline{c}}/{\rm d}y$ and ${\rm d}\sigma_{\rm b\overline{b}}/{\rm d}y$ at mid-rapidity in pp collisions at this collision energy. Furthermore, with the pp baseline and a new p--Pb measurement the $R_{\rm pPb}$ of dielectron productions as a function of \mee\ was measured, which does not suggest a modification of heavy-flavour production within uncertainties. The $R_{\rm AA}$ as a function of \mee\ on the other hand shows a hint for a modification in the IMR, most probably introduced by an interaction of the heavy quarks with the hot and dense medium created in AA collisions at LHC energies.
This complex interplay of initial and final state modifications of charm production in heavy--ion collisions complicates the construction of a precise baseline, making the measurement of thermal radiation of the medium or possible modifications due to partial restoration of chiral symmetry difficult. In future studies, this will be possible by separating the thermal radiation from off-vertex decays of heavy-flavour hadrons using their distinct decay topologies, an analysis which will benefit from the improved vertexing and impact parameter resolution of the inner tracking system upgrade for LHC Run3~\cite{itsUpgrade}.

%  \begin{itemize}
%      \item more PbPb statistics
%      \item better resolution with new ITS
%      \item faster data acquisition with new TPC readout
%  \end{itemize}



%% References


\bibliographystyle{elsarticle-num}
\bibliography{sample.bib}






% \end{thebibliography}

\end{document}